\documentclass[aps,prl,twocolumn,superscriptaddress,calc,epsfig,10pt]{revtex4-2}
\usepackage{graphicx}
\usepackage{graphics}
\usepackage{amssymb,amsmath}
\usepackage{comment}

\usepackage{wasysym}

\usepackage{color}
\usepackage{xcolor}
\usepackage{cancel}
\usepackage[normalem]{ulem}

\newcommand{\CUAaddress}{Harvard-MIT Center for Ultracold Atoms, Cambridge, Massachusetts 02138, USA}
\newcommand{\HarvardPhysicsAddress}{Department of Physics, Harvard University, Cambridge, Massachusetts 02138, USA}
\newcommand{\HarvardChemistryaddress}{Department of Chemistry and Chemical Biology, Harvard University, Cambridge, Massachusetts 02138, USA}

\newcommand{\kt}[1]{\ensuremath{\left|#1\right\rangle}}

\begin{document}

\title{Probing dipolar interactions between Rydberg atoms and ultracold polar molecules}

\author{Lingbang Zhu}
\thanks{These authors contributed equally to this work.}

\affiliation{\HarvardChemistryaddress}
\affiliation{\HarvardPhysicsAddress}
\affiliation{\CUAaddress}

\author{Jeshurun Luke}
\thanks{These authors contributed equally to this work.}

\affiliation{\HarvardChemistryaddress}
\affiliation{\HarvardPhysicsAddress}
\affiliation{\CUAaddress}

\author{Roy Shaham}
\affiliation{\HarvardChemistryaddress}
\affiliation{\HarvardPhysicsAddress}
\affiliation{\CUAaddress}

\author{Yi-Xiang~Liu}
\affiliation{\HarvardChemistryaddress}
\affiliation{\HarvardPhysicsAddress}
\affiliation{\CUAaddress}

\author{Kang-Kuen Ni}
\email{ni@chemistry.harvard.edu}
\affiliation{\HarvardChemistryaddress}
\affiliation{\HarvardPhysicsAddress}
\affiliation{\CUAaddress}
\date{\today}

\begin{abstract}
We probe resonant dipolar interactions between ultracold $^{40}$K$^{87}$Rb molecules and Rydberg $^{87}$Rb atoms in an optically trapped ensemble. Through state-selective ionization detection of the KRb molecules, we observe resonant energy transfer at 2.227~GHz from Rydberg atoms to molecules under a tunable external electric field. We measure a broadening up to 3.5~MHz, for the Rb Rydberg excitation spectrum, which matches a Monte Carlo simulation that describes a Rydberg atom and neighboring molecules evolving under a dipole-dipole interacting Hamiltonian. The demonstrated interspecies dipolar interaction is a key ingredient for hybrid Rydberg-polar molecule systems, where the advantages of each system can be leveraged and combined.

\end{abstract}
\maketitle
Controlled interactions between quantum systems facilitate energy exchange and entanglement, which form the basis for many demonstrations in quantum simulation and information processing. Most efforts to date have focused on perfecting individual physical platforms, leading to many advances. Combining different physical platforms with unique properties, such as their coherence time, electromagnetic spectrum coverage, and interactions~\cite{xiang2013hybrid,kurizki2015quantum,reiserer2015cavity,singh2023mid}, can expand the realm of quantum applications, offer improved performance, and mitigate the limitations of a single physical platform. Cold atoms and molecules are one of the leading platforms as both have been used to demonstrate gates~\cite{evered2023high,tsai2025benchmarking,picard2025entanglement}, simulate Hamiltonians~\cite{xu2023frustration,miller2024two}, and prepare quantum phases~\cite{choi2016exploring,de2019observation,norcia2021two,bigagli2024observation}. Hybrid systems of polar molecules and Rydberg atoms present an intriguing avenue combining many coherent degrees of freedom of molecules~\cite{park2017second,kondov2019molecular,burchesky2021rotational,ruttley2025long} and strong dipolar interactions of Rydberg atoms whose dipole moments ($d_a$) can reach $10^3-10^4$ times larger than those ($d_m$) of molecules~\cite{gallagher1994rydberg}.

Interspecies dipolar coupling ($\propto d_{a} d_{m}/r^3$, where $r$ is the separation of the two species) is central to hybrid Rydberg-polar molecule systems and has motivated many theoretical proposals. These include coherent control and non-destructive detection of polar molecules~\cite{kuznetsova2016rydberg, zeppenfeld2017nondestructive}, sympathetic and laser cooling of molecules~\cite{zhao2012atomic,huber2012dipole,zhang2024sympathetic}, simulation of spin-spin interactions~\cite{kuznetsova2018effective} and central spin model~\cite{dobrzyniecki2023quantum}, and fast high-fidelity quantum gates~\cite{zhang2022quantum,wang2022enriching}. Experimentally, resonant dipolar energy transfer from Rydberg atoms to polar molecules has been investigated in free-moving beams~\cite{petitjean1986collisions,gawlas2019rydberg, patsch2022rydberg, zou2022probing}. In trapped systems, Rydberg blockade from non-resonant interactions between a Rydberg atom and a polar molecule has been achieved~\cite{guttridge2023observation}. Combining these elements to realize resonant dipolar interactions in trapped ultracold atoms and molecules is a crucial step to unlocking the aforementioned applications. 

\begin{figure*}[t]
\centering
\includegraphics[width = 0.72\textwidth]{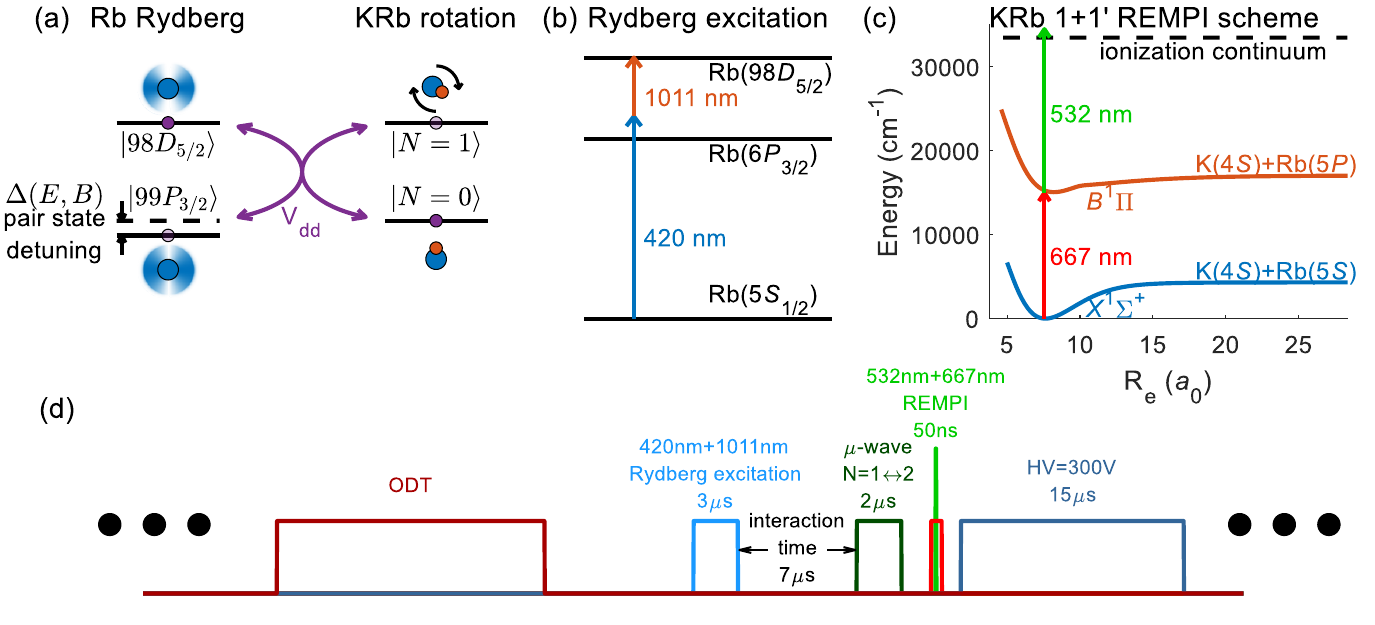}
\caption{Probing the resonant interaction between Rydberg Rb atoms and KRb molecules. (a) The resonant interaction is established between the atomic states $98D_{5/2}$ and $99P_{3/2}$, matched to the molecular rotational transition $N = 0 \rightarrow N = 1$. The pair state detuning $\Delta$, defined as $\Delta=\Delta E_{a}(98D_{5/2}-99P_{3/2})-\Delta E_m(N=1-N=0)$, can be tuned with an external electric and magnetic field. (b) Two-photon excitation scheme of Rb atoms to the $98 D_{5/2}$ state via the $6 P_{3/2}$ intermediate state, with two lasers at 420 nm and 1011 nm. (c) Detection scheme of the KRb rotation state. We use a 4.4 GHz microwave pulse to drive $N=1$ molecules to the $N=2$ state before ionization. The REMPI step consists of a 667 nm laser pulse transferring the KRb molecules from $X^1\Sigma^+$, $\nu=0$, $N$ = 2 to the electronic excited state $B^1\Pi,\nu'=0, N' = 1$, and a 532 nm laser pulse ionizing the molecules. Compared to directly ionizing $N=1$ molecules, this scheme reduced off-resonant ionization of background $N=0$ KRb. (d) Experimental sequence for a single excitation-detection cycle. In each cycle, we turned off the ODT and applied the Rydberg excitation pulse for 3 $\mu$s. The Rydberg atoms and KRb molecules interacted for 7 $ \mu $s before we detected the products. The timings were chosen to ensure the stabilization of Rydberg excitation and sufficient build-up of products. Following the microwave and REMPI pulses, we increased the voltage on the in-vacuum electrodes to ionize the Rydberg atoms through field ionization and extract both the Rb and KRb ions. }
\label{fig1}
\end{figure*}

\begin{figure}[h]
\centering
\includegraphics[width = 0.45\textwidth]{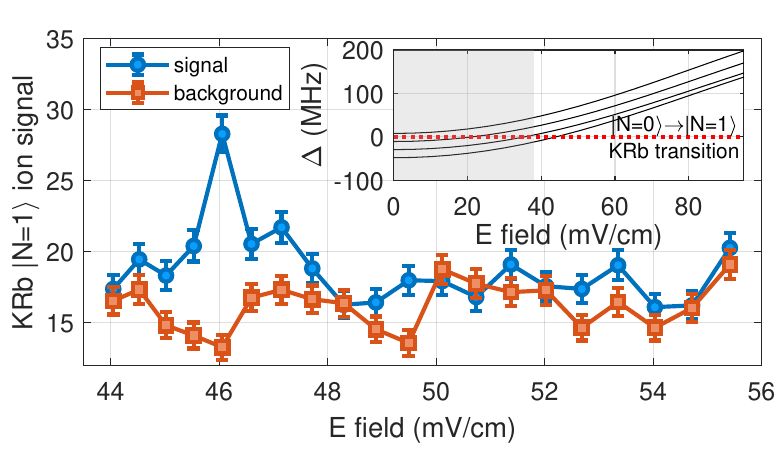}
\caption{Electric field tuning of resonant interactions between molecules and Rydberg atoms.  $N=1$  molecules are measured using REMPI at $B$ = 10 G. The data (blue circles) were taken with the Rydberg excitation pulse on resonance, and the background (orange squares) was measured with the pulse 70~MHz blue-detuned to the two-photon $5S_{1/2}$ to $98D_{5/2}$ transitions. No Rydberg excitation was observed for the 70~MHz detuned case. All error bars represent shot noise. (Inset) Calculated electric field dependence of the pair state detuning $\Delta$  for four possible transitions from the initial $98 D_{5/2}$ state to the $99 P_{3/2}, m_{J}=\pm\frac{3}{2}, \pm\frac{1}{2}$ states at 10 G. The dashed line represents the resonance condition. The shaded region is inaccessible experimentally.
}
\label{fig2}
\end{figure}

In this work, we experimentally demonstrated resonant dipolar interactions between Rydberg Rb atoms and KRb molecules in an optical dipole trap (ODT) and developed a quantitative model to explain our observations. We chose a pair of Rb Rydberg states that closely matches the energy difference (2.227~GHz) of the rotational ground and the first excited states of KRb (Fig.~\ref{fig1}(a)). We observed resonant energy transfer between the atoms and the molecules by directly detecting KRb in the rotational excited state. We tuned the Rydberg atoms and KRb molecules into resonance using external electric and magnetic fields and measured the broadening of the Rydberg excitation spectrum due to interspecies interactions. To understand and quantify this broadening, we modeled the Rydberg atom-molecule interactions with a dipole-dipole Hamiltonian and performed a Monte Carlo simulation.  We found good agreements between the computed excitation spectra and the experimental observations. 

We began our experiment by creating a mixture of $\mathrm{^{40}K^{87}Rb}$ molecules and $\mathrm{^{87}Rb}$ atoms at a temperature of $\mathrm{\sim400\ nK}$ following methods described in \cite{ni2008high,nichols2022detection}. The molecules were prepared in a single hyperfine state of the rovibronic ground state, and the Rb atoms were in the lowest hyperfine state $\kt{F=1,m_{F}=1}$ of the electronic ground state.
The atom-molecule mixture was confined in a crossed-beam ODT formed by two $1064$~nm laser beams. Following the creation of the mixture, the magnetic field was ramped down from 543.5 to 10 G in 10 ms. We allowed the field to stabilize for an additional 40~ms before probing the system for interspecies interactions.

We initiated the interactions between atoms and molecules with a Rydberg excitation pulse and probed the outcomes via either molecule or Rydberg atom detections, with relevant schemes and timings shown in Fig.~\ref{fig1}. The ODT was switched off to keep the atom-molecule cloud in the dark during the 3 $\mu$s excitation pulse. The excitation lasers drove the Rb atoms from the ground state to the $98D_{5/2}$ Rydberg state with a two-photon transition. Because the $98D_{5/2}$ state has a Rydberg blockade radius ($\sim$ 20 $\mu$m)  comparable to the cloud size, only a few Rydberg atoms were excited per pulse, and their spatial distribution was sparse. 

We probed the outcome of the interaction where we expect Rydberg atoms to exchange an energy quanta with nearby $N=0$ molecules as the appearance of $N=1$ molecules (Fig.~\ref{fig2}), which were ionized and detected through a state-selective 1+1' resonance-enhanced multi-photon ionization (REMPI) scheme~\cite{liu2024hyperfine}. Shortly afterward, we accelerated the KRb ions with an external electric field onto a micro-channel plate (MCP) to be counted. The residual Rydberg atoms were removed through field ionization. The excitation-detection step was repeated 5000 times over 1 second for each experimental cycle to allow the accumulation of statistics. 

To demonstrate the resonant nature of the Rydberg atom-molecule interactions, we tuned the pair state detuning $\Delta=\Delta E_{a}(98D_{5/2}- 99P_{3/2})-\Delta E_m(N = 1 - N = 0 )$ by varying the external electric field (Fig.~\ref{fig2}). This can be achieved because of the disparate dipole moments, $d_a=20$ kilo-Debye for atoms and $d_m=0.57$ Debye for molecules~\cite{ni2008high}, leading to a large differential DC Stark shift between the $98D_{5/2}$ and $99P_{3/2}$ states and a negligible differential shift for the molecular states. As we varied the electric field, the excitation laser frequency was tuned accordingly to maintain the two-photon resonance condition to the highest energy state in the $98D_{5/2}$ manifold. This excited state was chosen because it is the only state with resonances above the minimum achievable electric field at an applied magnetic field $B$ = 10 G. 

\begin{figure*}[t]
\centering
\includegraphics[width = \textwidth]{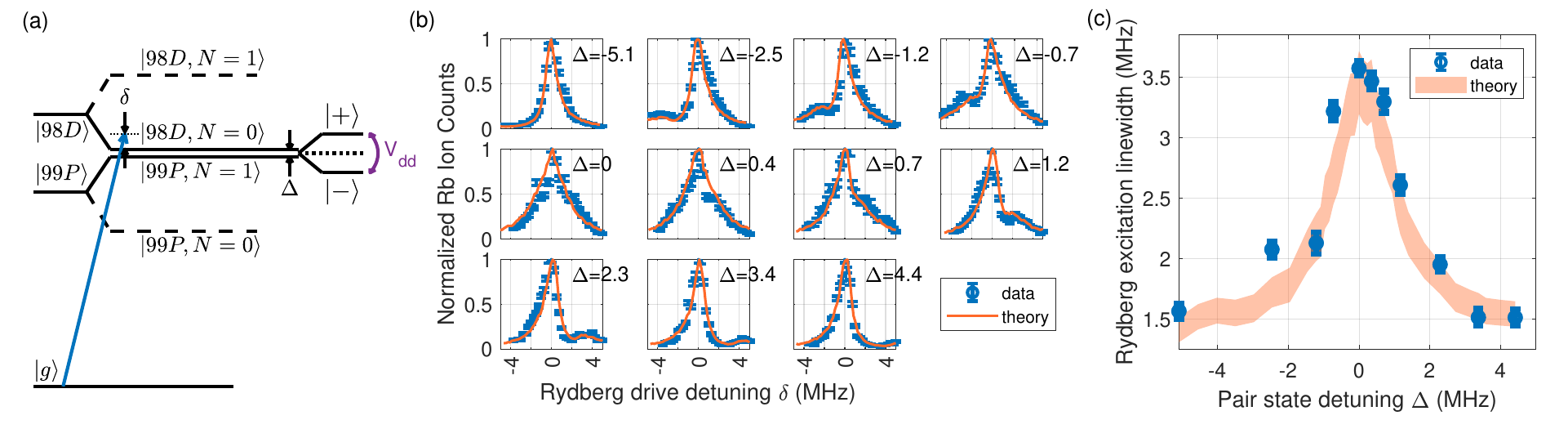}
\caption{Broadening of Rydberg excitation linewidth due to the Rydberg atom-molecule interactions. (a) Dressed state picture of an interacting Rydberg Rb atom and KRb molecule system. The eigenstates $|+\rangle$ and $|-\rangle$ are linear combinations of the bare states and are functions of the pair state detuning $\Delta$ and the interaction strength $V_{dd}$. $\delta$ represents the Rydberg drive detuning. (b) The experimental (blue) and theoretically calculated Rydberg excitation spectra (orange) for all pair state detuning $\Delta$ in units of MHz. The theoretical spectra are simulated using a Monte Carlo method based on a dipolar interaction model (SM Section B). (c) The FWHM linewidth of the Rydberg excitation transition as a function of $\Delta$. The error bars on the experimental data and the bounds of the orange band represent the 1-$\sigma$ confidence intervals. The experimental uncertainties account for shot noise in the ion signal, while the theoretical uncertainties include both Monte Carlo sampling error and uncertainties in the input parameters. The peak Rb and KRb densities are $\mathrm{4.29 \pm 0.36\ \times 10^{11}/cm^{3}}$ and $\mathrm{1.59 \pm 0.25\ \times 10^{10}/cm^{3}}$, respectively.  }
\label{fig3}
\end{figure*}

We note the application of the electric field and its calibration were deliberate. First, we characterized the residual electric field in the system via Rydberg spectroscopy (see SM Section A), which we found to be 96.5 $\mathrm{mV/cm}$. A small bias voltage was applied across the plate electrodes to cancel the residual field's projection along the normal axis of the plates, reducing the electric field to a minimum of 38.6 $\mathrm{mV/cm}$. 

As shown in Fig.~\ref{fig2}, we observed a narrow and prominent KRb ion signal at $E =\mathrm{46\ mV/cm}$. We identified this peak, which is close to the calculated resonance based on \cite{vsibalic2017arc}, to be the transition to the highest energy state in the $99P_{3/2}$ manifold, as shown by the intersection between the red dashed line and the bottom black curve in Fig.~\ref{fig2} inset. No additional resonances were detected at higher electric fields, which is consistent with theoretical predictions.

While ionization detection of $N=1$ KRb molecules provided a direct proof of the resonant energy transfer, we further probed the underlying interspecies interaction that manifested in the Rydberg excitation spectrum. Intuitively, near resonance ($\Delta \lesssim V_{dd} $), the interaction $V_{dd}$ between a Rydberg atom and a molecule couples the states $\{|98D, N = 0\rangle,|99P, N = 1\rangle\}$, forming two dressed states $|+\rangle$ and $|-\rangle$ as illustrated in Fig.~\ref{fig3}(a), both of which can be connected by the Rydberg drive from the ground state. This would lead to a double-peak excitation spectrum, where $V_{dd}$ set the spacing between the two peaks. However, due to the random spatial distribution of the atoms and molecules in the mixture, $V_{dd}$ varies, resulting in an inhomogeneous broadening of the Rydberg excitation.

We measured the broadening of the excitation spectrum for each pair state detuning $\Delta$, as shown in Fig.~\ref{fig3}(b). Each spectrum is fitted to a Lorentzian lineshape as an approximation to extract a full-width-half-max (FWHM) linewidth for comparison in Fig.~\ref{fig3}(c). When the Rydberg atom and KRb were on resonance, the linewidth (3.5 MHz) was significantly broader than the off-resonant case (1.5 MHz), where the off-resonant linewidth was primarily limited by decoherence during the Rydberg excitation.

To provide a theoretical model for our observed Rydberg excitation spectra (see SM Section B), we treated both the Rydberg atom and the molecule as point dipoles that interact under an electric dipole-dipole Hamiltonian \( H_{DD} = \frac{1}{4 \pi \epsilon_0 } \frac{\mathbf{d}_{\mathrm{a}}\cdot \mathbf{d}_{\mathrm{m}} - 3 (\mathbf{d}_\mathrm{a} \cdot {\hat{r}})(\mathbf{d}_\mathrm{m} \cdot {\hat{r}})}{r^3} \), where $\mathbf{d}_{\mathrm{a}}$, $\mathbf{d}_{\mathrm{m}}$ are the dipole operators of the Rydberg atom and molecule respectively, and $\hat{r}$ is the spatial separation unit vector between the species. The distributions of separations and relative orientations between Rydberg atoms and surrounding KRb molecules are accounted for through a Monte Carlo simulation. For each Rydberg atom, we included the 10 nearest KRb molecules in the system by considering the pairwise interaction between the Rydberg atom and every molecule (see SM Section B for details). We found that including at least three nearest KRb molecules is sufficient to reproduce nearly 90\% of the observed excitation linewidth, and incorporating additional neighbors (which are farther away) leads to a saturation of the simulated broadening, in good agreement with experimental observations (see SM Section B). Therefore,  in additional to inhomogeneous broadening, collective enhancement is included in the model to capture the entirety of the broadening in the observations.

The simulated Rydberg spectra are overlaid in Fig.~\ref{fig3}(b) for each $\Delta$ and agree well with the measurements. For a comparison of the excitation linewidths, we also fit the simulated spectra to a Lorentzian lineshape to extract FWHM, shown as the orange band in Fig.~\ref{fig3}(c). Overall, our theoretical calculations reproduced the broadened excitation spectra. We note that even though our theoretical model assumed that the resonant interaction is coherent, the ensemble averaging process washed out oscillations in the Rydberg spectroscopy data, and separately, time evolution measurements of the $N=1$ KRb molecule population.

Because the dipolar interaction strength $V_{dd}$ depends on the distance between the Rydberg atom and the nearest molecule, a reduction in spectral broadening is expected at lower KRb densities. Additionally, as the separation between Rb atoms and neighboring KRb molecules increases, the collective enhancement effect diminishes, further contributing to the decrease in spectral broadening. To verify this, we controllably reduced the KRb density. The experimental implementation (SM Section A) started with exciting a desired portion of the KRb molecules to $N=1$ via a microwave pulse, followed by dissociating the remaining $N=0$ molecules into atoms. We then removed all K atoms and part of the Rb atoms with resonant light. Finally, we used a microwave $\pi$-pulse to transfer the hide-out $N=1$ population back to $N=0$.

\begin{figure}[t]
\centering
\includegraphics[width = 0.47\textwidth]{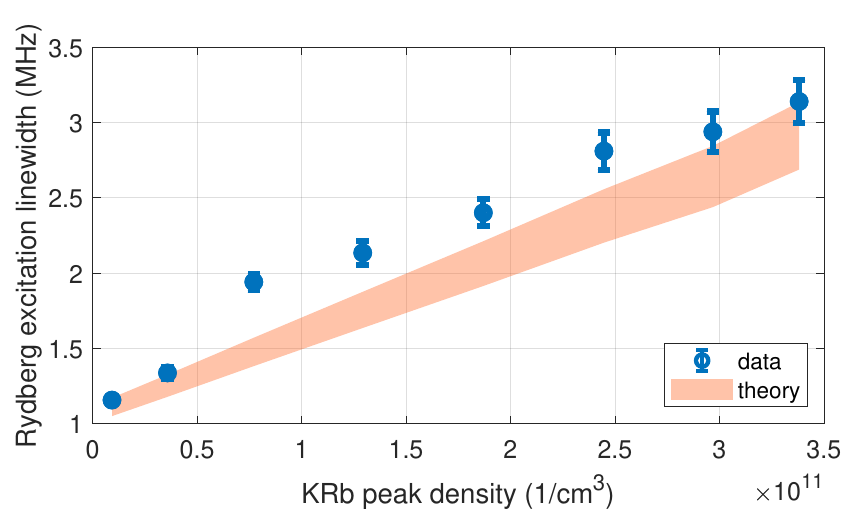}
\caption{Rydberg excitation linewidth as a function of peak KRb molecule density. We varied the KRb density up to $\mathrm{3.38 \times 10^{11}/cm^{3}}$. The experimental excitation linewidths were extracted from the FWHM of Lorentzian fits of the measured spectra. Theoretical linewidths were obtained by applying the same fitting procedure to the simulated spectra, given the KRb densities and relevant experimental parameters. The KRb density remained relatively constant during the Rb atom excitation and depletion, with less than a 10\% decrease observed over this interval. }
\label{fig4}
\end{figure}

We extracted the FWHM of the Rydberg excitation spectrum for each KRb density, as shown in Fig.~\ref{fig4}. The excitation linewidths broadened from 1.1 MHz to 3.2 MHz as we increased the KRb density from 0 to $\mathrm{3.38 \times 10^{11}/cm^{3}}$, and the theoretical calculations presented a good match with experiment data. We independently verified that varying the Rb density did not affect the excitation linewidths (SM Section A), as the Rydberg atom density remained unchanged due to the Rydberg blockade effect.

While the dipole-dipole interaction model captures the experimental results, the point-dipole approximation is expected to break down when the interspecies separation becomes comparable to the size of the Rydberg orbital. A charge-dipole interaction Hamiltonian, separating contributions of the valence electron and the positively charged core, could deviate from dipole-dipole interaction at higher KRb densities. Future works can test for the applicability of such a Hamiltonian.

This work provides an experimental demonstration and measurement of resonant dipolar interactions between ultracold Rydberg atoms and polar molecules, paving the way for realizing hybrid systems for quantum computation and simulation. A potential next step is to explore the central spin model at higher molecular densities, where one may also investigate the physics of a strongly interacting molecular gas mediated by Rydberg atoms. In parallel, coherent interaction dynamics could be more naturally realized in a system with uniform particle spacing, such as individually optical-tweezer-confined atoms and molecules, aiding non-destructive detection of molecules and speeding up quantum gates between molecules~\cite{zhang2022quantum,wang2022enriching}.

\section*{Acknowledgment}
We thank Grigor Adamyan and Artem Volosniev for fruitful discussions of charge-dipole interaction. We thank Rosario González-Férez, Mikhail Lemeshko, and David Wellnitz for discussions. We thank Mark Babin for experimental assistance. This work is supported by the U.S. Department of Energy (DOE), Office of Science, Basic Energy Sciences (BES), under Award No. DE-SC0024087 (molecule state detection), the Center for Ultracold Atoms (an NSF Physics Frontiers Center,  PHY-2317134) (Rydberg excitation), AFOSR DURIP FA9550-23-1-0122 (instrument upgrade), and the Gordon and Betty Moore Foundation GBMF11558 (Theoretical calculations). 

\bibliographystyle{apsrev4-2}

\onecolumngrid
\newpage
\section{Supplementary Materials Section A: Experimental Methods}

\subsection{Electric field calibration}

We calibrated the electric field using Rydberg spectroscopy. The electrode setup consists of six concentric plate electrodes spaced along the time-of-flight axis. During Rydberg excitation, we applied an external electric field by setting a controllable voltage $V_{\mathrm{rep}}$ on the first plate electrode (the repeller). The electrodes were connected through a voltage divider network, such that the voltages on the six electrodes follow a ratio of $1:0.82:0.585:0.35:0.175:0$. In addition, a residual electric field $E_{\mathrm{residual}}$ existed in the system even when all electrodes were grounded. Therefore, it is essential to calibrate the total electric field as a function of the applied repeller voltage to ensure accurate field control.

\begin{figure}[h]
\centering
\includegraphics[width = 0.8\textwidth]{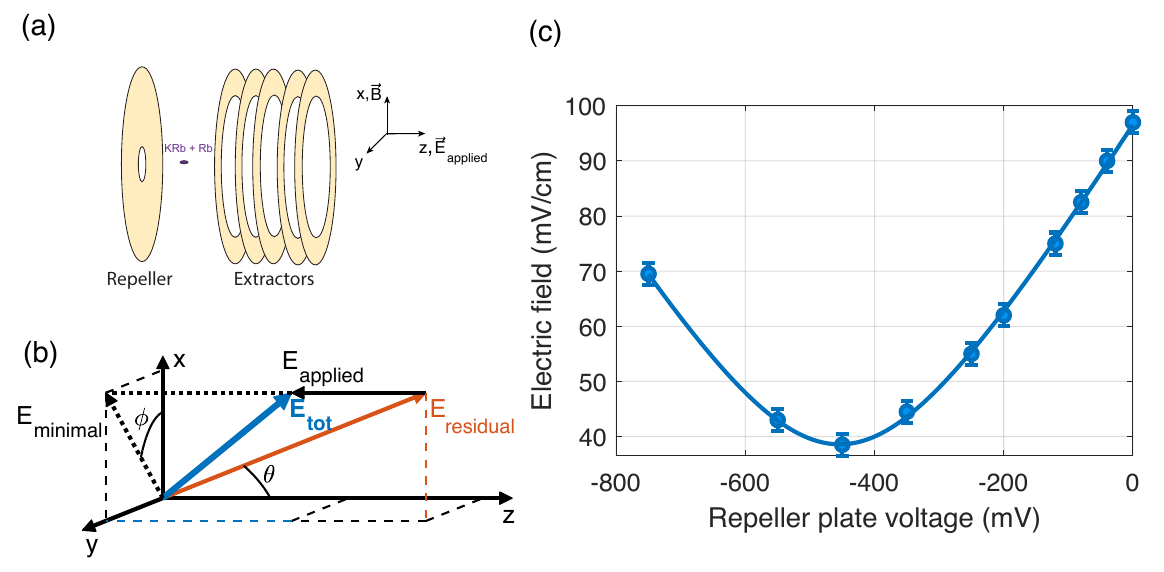}
\caption{Geometry and calibration of the electric field. (a) Field plates used to apply the appropriate E-field. The leftmost plate is the repeller, and the other plates on the right are extractors.  The Rb-KRb ensemble is placed at the center of the repeller and the first extractor plate. (b) Geometry of the electric field in the system. $E_{\mathrm{residual}}$ is the residual electric field when all plate electrodes are grounded, which corresponds to a $E_{\mathrm{{applied}}}$ of zero. We can completely cancel the residual electric field in the z-direction, after which the total electric field would be minimized to $E_{\mathrm{minimal}}$. (c) Calibration of the residual electric field by varying the repeller plate voltage. The electric field is extracted from comparing $98 D$ Rydberg states excitation spectrum at zero magnetic field to theoretical calculations. The blue curve is fitted based on Eq.~\ref{eq:E_calibration}. }
\label{E_field_compensation}
\end{figure}

The electric field geometry is shown in Fig.~\ref{E_field_compensation}(a/b). Our goal is to measure all 3 projections of the residual electric field $E_{\mathrm{residual}}$. Here, the x-axis is the magnetic field direction, and the z-axis is the symmetry axis of the plate electrodes. Due to the concentric geometry of the electrodes, the applied electric field only has a z-component. The total electric field magnitude with an applied voltage $V_{\mathrm{rep}}$ is:

\begin{align}
    E_{tot} = \sqrt{E_{minimal}^{2} + (E_{res,z} + \alpha V_{rep})^{2}}.
\label{eq:E_calibration}
\end{align}
$E_{\mathrm{minimal}}$ and $E_{\mathrm{res, z}}$ are the perpendicular and parallel components of the residual electric field relative to the z-axis. $\alpha$ is a conversion factor between the applied voltage and the electric field $E_{\mathrm{applied}}$. To measure these quantities, we adjusted the repeller voltage $V_{\mathrm{rep}}$ in the absence of a magnetic field. At each voltage, we performed Rydberg spectroscopy scans for the $98D$ states, which were compared with theoretical calculations to extract the electric fields as shown in Fig.~\ref{E_field_compensation}.

After fitting total electric fields at different applied voltages to Eq.~\ref{eq:E_calibration}, we determined $E_{\mathrm{minimal}}$ = 38.6 mV/cm, $E_{\mathrm{res, z}}$ = 88.5 mV/cm, and $\alpha$ = 0.1953 (mV/cm)/mV. In addition to this, we also wish to know the angle $\phi$ between the x-axis and $E_{\mathrm{minimal}}$, as they also affect the compositions and energies of the eigenstates under a finite magnetic field. This was achieved by measuring the Rydberg excitation spectrum at a magnetic field of $B$ = 10 G and comparing it with theoretical predictions, where we determined $\phi=55^\circ$. Using our simulation, we also found that $\phi$ had little impact on the Rydberg excitation broadening.

\subsection{Controlling KRb density with a microwave transfer scheme}

\begin{figure}[h]
\centering
\includegraphics[width = 0.45\textwidth]{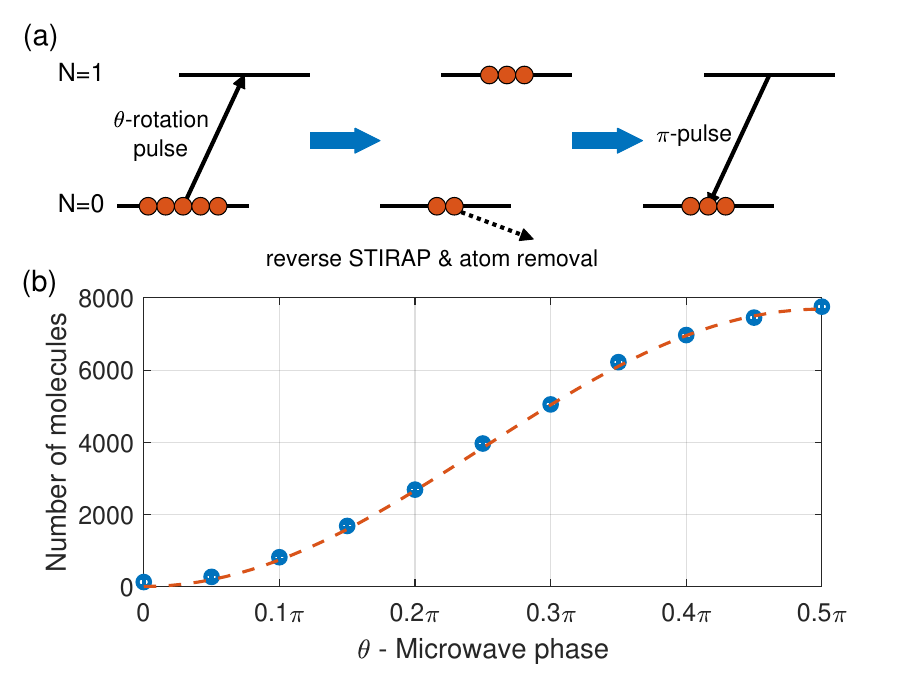}
\caption{Microwave sequence to control the KRb number density. (a) Shelving sequence to partially remove KRb molecules. Ground state molecules were first partially driven to the first rotational excited state. The remaining molecules in the ground state were then removed, after which we transferred the excited molecules back to the ground state. (b) The remaining molecule number vs the rotation angle $\theta$ on the Bloch sphere. The result is fitted to a sinusoidal function, as expected for a Rabi oscillation. The number of returning molecules is extracted from absorption imaging, with error bars reflecting uncertainty in the number of imaged molecules. }
\label{fig:uwave}
\end{figure}

We used a microwave shelving scheme to control the KRb density by partially removing KRb molecules from the cloud, as shown in Fig.~\ref{fig:uwave}(a). We started with KRb molecules in the ground rotation state. A microwave pulse drove the molecules to a superposition state between $|N=0\rangle$ and $|N=1\rangle$, corresponding to a $\theta$ rotation on the Bloch sphere. We removed the $|N=0\rangle$ KRb molecules using a reversed STIRAP pulse to drive the molecules to weakly-bound Feshbach states. These weakly bound molecules, along with part of the Rb and all K atoms, were removed by resonant laser pulses. Through varying the duration of the first $\theta$-rotation pulse, we control the number of molecules in the ensemble between 0 and $7.69\times10^{3}$, corresponding to a peak molecular density between 0 and $\mathrm{3.38 \times 10^{11}/cm^{3}}$, as shown in Fig.~\ref{fig:uwave}(b).

\subsection{Effect of atom density on the Rydberg excitation linewidth}

\begin{figure}[h]
\centering
\includegraphics[width = 0.45\textwidth]{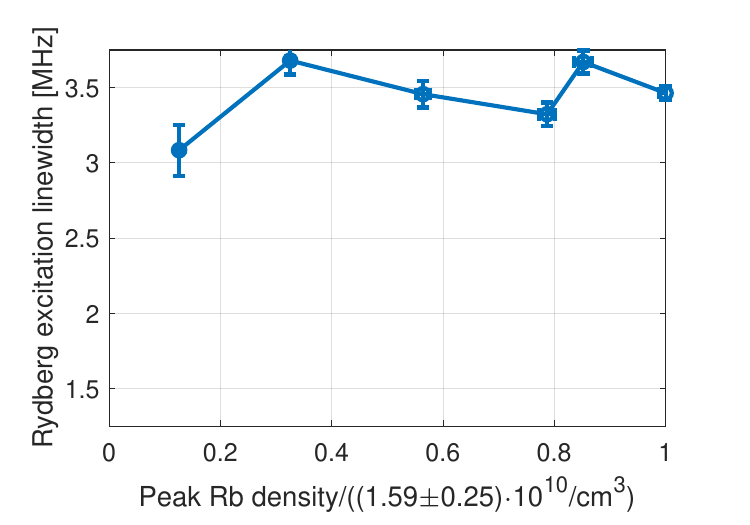}
\caption{Rydberg excitation linewidth versus Rb atom density. We fitted the Rydberg excitation spectra to a Lorentzian function and extracted their FWHM linewidths using the same method as in Fig.~\ref{fig3} and Fig.~\ref{fig4} of the main text. The KRb density was kept at the highest achievable value of $\mathrm{4.29 \times 10^{11}/cm^{3}}$, identical to conditions used in main text Fig.~\ref{fig3}. }
\label{atom_density}
\end{figure}

Because the resonant interaction is between Rydberg atoms and KRb molecules, the Rydberg excitation linewidth should not depend on the Rb atom density in the low atom density regime ($\mathrm{\sim 10^{10}/cm^{3}}$) \cite{liebisch2016controlling}. We experimentally tested this by measuring the Rydberg excitation linewidths at different Rb atom densities. To do this, we removed part of the Rb atoms inside the cloud with a microwave adiabatic rapid passage (ARP) pulse that transferred these Rb atoms to the $\kt{F=2,m_{F}=2}$ state, which was subsequently expelled by a resonant laser pulse. Meanwhile, we kept the highest achievable KRb density of $\mathrm{4.29 \times 10^{11}/cm^{3}}$ and the atoms and molecules on resonance as we varied the Rb density. Overall, we found that the Rydberg excitation linewidth does not depend on the Rb density, as shown in Fig.~\ref{atom_density}.

\section{Supplementary Materials Section B: Theory of the interaction dynamics}

\subsection{Dynamics under the dipole-dipole Hamiltonian }

We modeled the energy exchange between a Rydberg atom and a KRb molecule by treating both species as point dipoles interacting under the dipole-dipole Hamiltonian. We evaluated the operators over a basis composed of the tensor product of the subspaces for the molecule and the atom. The molecular subspace was spanned by the rigid rotor eigenstates $\{|N, M_N\rangle\ (N = 0, 1)\ (-N\le M_N\le N)\} $. The atomic subspace was spanned by the basis $\{|L, m_L, S = 1/2, m_S \rangle \ (L = 1, 2)\ (-L\le m_L\le L)\ (m_S =\pm1/2)\}$. Within this basis, the matrix element of the dipole operator for the molecule can be evaluated with, 
\begin{equation}
\begin{split}
    \langle N' M_{N'}|d^{q}_{\mathrm{m}}| N M_N\rangle =& d_0 (-1)^{M_{N'}} \sqrt{(2N' + 1)(2N + 1)} \\
&\begin{pmatrix}
N' & 1 & N \\
-M_{N'} & q & M_N
\end{pmatrix} 
\begin{pmatrix}
N' & 1 & N \\
0 & 0 & 0
\end{pmatrix}, 
\end{split}
\end{equation}
where $d_0$, the permanent dipole moment of KRb molecules, is 0.57 Debye \cite{ni2008high}. Similarly, the atomic dipole operator in the coupled basis $| n, J, M_J, L, S\rangle$ can be evaluated as~\cite{zare1988angular}, 
\begin{equation}
    \langle n', J', M_J' , L', S' | d^{q}_{\mathrm{a}} | n, J, M_J, L, S\rangle  = e \delta_{S S'} (-1)^{L' + S' + J + 1} \sqrt{(2J + 1)(2J'+ 1)} \begin{Bmatrix}
        L' & J' & S' \\
        J & L & 1
    \end{Bmatrix} \langle n', L', J|| d_{\mathrm{a} }|| n, L, J \rangle
\end{equation}
with, 
\begin{equation}
     \langle n', L', J'|| d_{\mathrm{a} }|| n, L, J \rangle = (-1)^{L'} \sqrt{(2L + 1)(2L' + 1)}\begin{pmatrix}
         L' & 1 & L \\ 0 & 0 & 0 
     \end{pmatrix} \int_0^{\infty}R_{n', L', J'}(r)rR_{n, L, J}(r)r^2 dr
\end{equation}
where $R_{n, L}$ is the radial wavefunction for the Rydberg state, and q ranges from -1 to 1, corresponding to the different spherical components of the dipole vector operator. Given that we are interested in operators represented by the uncoupled basis, we use the appropriate Clebsch-Gordan coefficients to go from the coupled to the uncoupled basis. 

In addition to the dipole-dipole coupling, the total Hamiltonian also includes the individual molecule and atom terms, 
\begin{equation}
    H_0 = H_{DD}  +H_{\mathrm{atom}}  + H_{\mathrm{mol}}.
    \label{eq:H0}
\end{equation}
Because rotation is the only relevant degree of freedom for the molecules, 
\begin{equation}
    H_{\mathrm{mol}} = B \mathbf{N}^2 = 2\pi\times2.227\,\mathrm{GHz}\times|N=1\rangle\langle N=1|,
\end{equation}
where $B$ is the rotational constant of KRb molecules~\cite{ni2008high}, and the second equality is evaluated in the chosen molecular subspace. Within this approximation, we ignored the hyperfine structure within each rotational state as their interaction strengths are on the order of 0.1~MHz, and are much weaker than the average dipolar interactions between Rydberg atoms and molecules. Therefore, we assumed different $M_N$ projections in the $N = 1$ manifold to be energetically degenerate in the following calculations.

The atomic Hamiltonian includes the Coulombic, spin-orbit, Stark, and Zeeman terms, all of which were explicitly included in the calculations because their energy scales are greater than the dipolar coupling.
We used the ARC Rydberg package \cite{vsibalic2017arc} to calculate the Rydberg atom eigenstate energies and decompositions by diagonalizing the Hamiltonian in the coupled basis with $n$ ranging from 96 to 100 and $L$ from 0 to 10, respectively. We used this larger subspace for the atom Hamiltonian to reproduce the spectroscopic results accurately. However, this large subspace was too big to compute the dipolar interaction time dynamics due to the computational complexity.

To reduce the Hamiltonian to a smaller and more tractable subspace ($99P_{3/2}$, $98D_{5/2}$), we truncated the eigenstates $U^{full}$ in the original subspace, resulting in the matrix $U$. Because this new matrix is no longer orthonormal, we decomposed our truncated unitary with an SVD decomposition $U = M \Sigma V$ and imposed singular values of unity to arrive at an orthonormalized basis set $U' = MV$. Under this approach, the atomic state that resonantly interacts with the molecule maintains its overlap with the original eigenstates after orthonormalization. The truncated atomic Hamiltonian in the uncoupled basis can be calculated using the orthonormalized wavefunctions with, 
\begin{equation}
    H_{\mathrm{atom}} = U' H^{full}_{\mathrm{atom}} (U')^{\dagger},
\end{equation}
where $H^{full}_{\mathrm{atom}}$ is the diagonal matrix with energies associated with relevant eigenstates before truncation. This approach allowed us to capture the exact eigen-energies and the approximate eigenstate decompositions.

\subsection{Collective Interactions}

\begin{figure}[h]
\centering
\includegraphics[width = 0.45\textwidth]{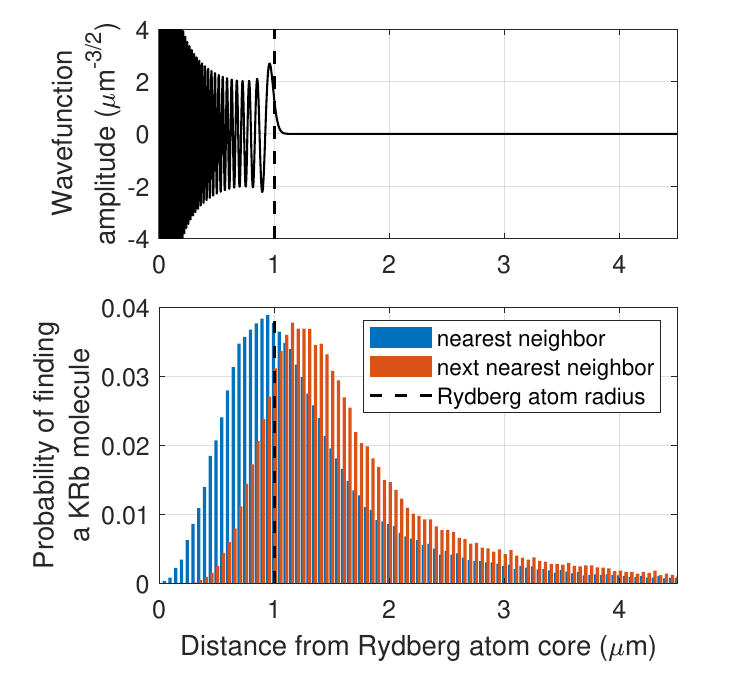}
\caption{Wavefunction of the $98D_{5/2}$ Rydberg electron and the averaged distance distribution of the nearest and the next nearest KRb molecules from a Rydberg atom. We determined the Rydberg atom radius to be 1 $\mu$m (dashed line) as the wavefunction falls off dramatically beyond it. This corresponds to a characteristic KRb density of $\mathrm{1 \times 10^{12}/cm^{3}}$. There is a 39\% probability of finding a single KRb molecule within the Rydberg orbit and a 14\% probability of finding two.}
\label{fig:nearest_neighbor}
\end{figure}

\begin{figure}
    \centering
    \includegraphics[width=0.4\linewidth]{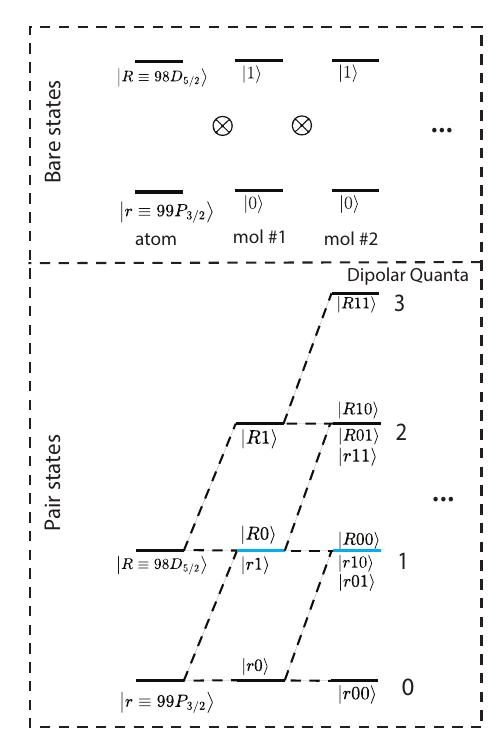}
    \caption{Bare atomic and molecular states used to form pair states. While the amount of states grows exponentially ($2^{n +1}$) with the number of molecules $n$, the number of pair states with only one dipolar quanta grows linearly as $n + 1$. }
    \label{fig:Dressed-States}
\end{figure}

We perform Monte Carlo simulations by sampling particle positions based on measured spatial dimensions and densities of the atomic and molecular clouds. The distribution of the distances between a random Rb atom and its nearest and next-nearest molecular neighbors are presented in Fig.~\ref{fig:nearest_neighbor}. A finite probability of having multiple KRb molecules inside the Rydberg atom wavefunction motivated the use of a collective model where one Rydberg atom interacts with several of KRb molecules. Because a collective model exhibits exponential scaling of the Hilbert space dimension with increasing molecule number, we restrict our calculations to the single-excitation subspace as further expanded upon in Fig.~\ref{fig:Dressed-States}. 

This approach ensures a linear scaling of the Hilbert space dimension, making the computations tractable. We can now define $H_0$ (Eq.\,\ref{eq:H0}) within this relevant subspace $\{|R\rangle |0, 0, 0 ...0\rangle$, $|r\rangle |1^m_j\rangle \}$, where $|1^m_j\rangle = | 0, 0,...,1^m, ..0\rangle$  denotes the excitation of the $j$-th molecule to one of the three $m= (1, 0, -1)$ molecular projection modes of $|N = 1, M_N = m\rangle$. This approximation is valid due to the conservation of energy, given that the interaction strength is sufficiently weaker than the energy of dipolar quanta (2.227 GHz). In addition, we also simplified our treatment of the Rydberg atom as a two-level system by focusing solely on the coupling between the highest energy $98D_{5/2}$ state ($|R\rangle$) and the highest energy $99P_{3/2}$ state ($|r\rangle$).
Interactions between other state pairs are at least 20 MHz detuned and, thus, are significantly suppressed, ensuring the closure of the chosen subspace and the validity of the approximation.

\begin{figure}[h]
\centering
\includegraphics[width = 0.45\textwidth]{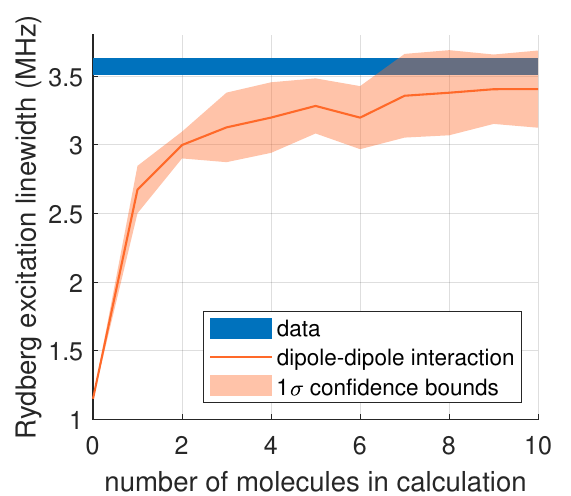}
\caption{Simulated spectral broadening versus the number of nearest KRb molecules in the Monte Carlo model. Blue: 1-$\sigma$ confidence interval of the measured broadening. Orange: 1-$\sigma$ confidence interval of the simulated Rydberg spectrum's FWHM as a function of the number of included KRb molecules ($n = 1$ to $10$). FWHM are extracted from fitting a Lorentzian curve to measured and calculated spectra. }
\label{fig: number of molecules in simulation}
\end{figure}

In a tractable simulation, we include up to 10 nearest KRb molecules per Rydberg atom. Fig.~\ref{fig: number of molecules in simulation} shows the FWHM of the simulated Rydberg spectra close to the pair state resonance as a function of the number of KRb molecules included in the calculations. The results indicate that incorporating at least three nearest neighbors is necessary to reproduce approximately 90\% of the observed spectral broadening. As we include more molecules, the simulated broadening saturates, suggesting diminishing contributions from more distant neighbors.

\subsection{Monte Carlo simulation}

Using the techniques discussed above, we can generate the Hamiltonian to describe the interaction of $n$ molecules with a Rydberg atom. The experimental observable we wish to capture is the Rydberg population as a function of the Rydberg drive detuning. As such, we expanded the Hamiltonian, previously defined only in the Rydberg/molecule subspace (Eq.\,\ref{eq:H0}), to the basis including the ground state of the atom  $\{|R\rangle |0, 0, 0 ...0\rangle$, $|r\rangle |1^m_j\rangle \}\oplus {|g\rangle |0, 0, 0, ..., 0\rangle} $ with, 
\begin{equation}
\begin{split}
    H =& H_0 \oplus (\delta |g\rangle |0, 0, 0, ..., 0\rangle \langle g | \langle 0, 0, 0, ..., 0| \\
    &+  \Omega |g\rangle |0, 0, 0, ..., 0\rangle \langle R| \langle 0, 0, 0, ..., 0| + \mathrm{h.c})
    , \end{split}
    \label{Eq:Hamiltonian_aton_mol}
\end{equation}
where $\delta $ is the drive detuning and $\Omega$ is the Rydberg drive Rabi rate. 

The time dynamics over the 3 $\mu s$ of the Rydberg drive was carried out using a master equation with the Hamiltonian in Eq.~\ref{Eq:Hamiltonian_aton_mol} and the dephasing Lindblad term generated by the operator $\sqrt{\gamma} |g\rangle\langle g|$ to capture drive decoherence from the phase noise of the laser. The Rabi rate of $\Omega = 2\pi \times 50$ kHz was selected from calculations based on beam intensity and polarization. Decoherence rate of $\mathrm{\gamma = 2\pi \times 900\ kHz}$ was selected to match the Rydberg excitation linewidth without resonant interactions with molecules as measured at large pair state detuning $\Delta$.
The initial state of the system was $|g\rangle|0,0,0,\ldots,0\rangle\langle g|\langle 0,0,0,\ldots,0|$.

In addition to the expansion of Hilbert space, we also needed to account for the inhomogeneous broadening present as a result of the randomized interaction distances in the ensemble. This could be realized with a Monte Carlo simulation of the experimental cycles by initializing molecules and atoms according to a 3-dimensional normal distribution set by the cloud dimensions directly measured in the experiment ($\mathrm{Rb}: (\sigma_x = 7.3 \pm0.25, \sigma_y = 53.2, \pm 4.3, \sigma_z = 7.3 \pm 0.25)\, \mu \mathrm{m} $; $\mathrm{KRb}: (\sigma_x = 6.0 \pm 0.34, \sigma_y = 36.3 \pm 1.7, \sigma_z =6.0 \pm 0.34)\, \mu \mathrm{m} $). For each cycle, we randomly picked one Rb atom from the cloud to be driven to the Rydberg state with a Rabi-rate accounting for the finite beam diameters of 60 $\mu m$ and 55 $\mu$m for the 1011 nm and 420 nm drive, respectively. The drive was simulated in the presence of dynamics of the Rydberg atom interacting with its 10 nearest KRb molecules. The Rydberg population from each cycle was extracted by taking the observation value of the density matrix, and it was averaged over 512 repetitions for a given set of experimental parameters.

\end{document}